# Transmission infrared micro-spectroscopic study of individual human hair

Chen Li[1*#], Yuhan Du[2#], Haonan Chen[3], Xinxin Han[4], Wenbin Wu[2], Xiufang Kong[5], Cheng Zhang[3,6*], Xiang Yuan[2,7*]

[1]Department of General Practice, Zhongshan Hospital, Fudan University, Shanghai, China

[2]State Key Laboratory of Precision Spectroscopy, East China Normal University, Shanghai, China

[3]State Key Laboratory of Surface Physics and Institute for Nanoelectronic Devices and Quantum Computing, Fudan University, Shanghai, China

[4]Department of General Practice, Peking Union Medical College Hospital, Chinese Academy of Medical Science and Peking Union Medical College, State Key Laboratory of Complex Severe and Rare Diseases, Beijing, China

[5]Department of Rheumatology, Zhongshan Hospital, Fudan University, Shanghai, China

[6]Zhangjiang Fudan International Innovation Center, Fudan University, Shanghai, China

[7]School of Physics and Electronic Science, East China Normal University, Shanghai, China

[*]Correspondence and requests for materials should be addressed to C. L. (E-mail: li.chen2@zs-hospital.sh.cn), C. Z. (E-mail: zhangcheng@fudan.edu.cn) & X. Y. (E-mail: xyuan@lps.ecnu.edu.cn)

[#] These authors contributed equally to this work

## Abstract

Understanding the optical transmission property of human hair, especially in the infrared regime, is vital in physical, clinical, and biomedical research. However, the majority of infrared spectroscopy on human hair is performed in the reflection mode, which only probes the absorptance of the surface layer. The direct transmission spectrum of individual hair without horizontal cut offers a rapid and non-destructive test of the hair cortex but is less investigated experimentally due to the small size and strong absorption of the hair. In this work, we conduct transmission infrared micro-spectroscopic study on individual human hair. By utilizing direct measurements of the transmission spectrum using a Fourier-transform infrared microscope, the human hair is found to display prominent band filtering behavior. The high spatial resolution of infrared micro-spectroscopy further allows the comparison among different regions of hair. In a case study of adult-onset Still's disease, the corresponding infrared transmission exhibits systematic variations of spectral weight as the disease evolves. The geometry effect of the internal hair structure is further quantified using the finite-element simulation. The results imply that the variation of spectral weight may relate to the disordered microscopic structure variation of the hair cortex during the inflammatory attack. Our work reveals the potential of hair infrared transmission spectrum in tracing the variation of hair cortex retrospectively.

**Keywords:** Transmission infrared micro-spectroscopy, Human hair, Spectral weight variation, Finite-element simulation

## Introduction

Infrared light refers to a wide range of electromagnetic radiation with a wavelength longer than that of visible light. As its spectra range covers the energy of vibrational modes of chemical bonds, infrared spectroscopy is widely recognized as a powerful approach for the identification of chemical components and functional materials in physical, chemical, and biomedical research.[1–4] Meanwhile, infrared light excites chemical bonds and evokes various photothermal and photobiological responses in tissues.[5–8] It increases local circulation and relieves inflammation, gradually becoming an emerging clinical rehabilitation tool known as infrared therapy nowadays.[9–11] In order to deliver proper light dose in the target, like the abdomen or brain, it is crucial to investigate and analyze the infrared spectra of cutaneous appendages such as nails and hair.

Hair is fiber-like, with a typical diameter of about 50-100 μm, and has strong absorption in the near and middle infrared (NIR and MIR) regimes.[12–15] As shown in Fig. 1a, hair consists of three parts: a thin protective cuticle layer, a middle cortex layer, and an innermost medulla, each showing a distinct microscopic structure.[14] The complex features pose a challenge to precisely measure the infrared spectrum of an individual hair. Early works focused mainly on performing Fourier-transform infrared spectroscopy (FTIR) measurements on a cluster of hairs to enhance the transmission signals.[13] Later, the absorption spectra of one single hair were studied intensively with the development of FTIR microscopy.[16–19] The infrared spectrum of hair is intimately connected with the helix and disorder of the keratin structure. Changes in spectral weight have been found in hairs that have undergone specific treatments such as bleaching, oxidation, and chemical diffusion.[18,20,21] The chemometric analysis on infrared spectra enables the identification of hair-like samples from different species as well as synthetic fibers. It provides a rapid methodology for the forensic examination of potential hair evidence.[22] Meanwhile, infrared spectra of hairs from a series of cancer patients suggest significant loss or change of hair medulla, which may contribute to the early diagnosis or cancer relapse in the future.[23–29]

However, most infrared spectroscopic study of hair has been performed in the attenuated total reflection (ATR) mode, which unfortunately probes the absorptance of the surface layer only. Detecting internal parts in hair in ATR-FTIR requires sophisticated microtomy, which may introduce external contamination to the infrared spectrum measurement. As Fig. 1b shows, ATR probes the sample information by the evanescent wave generated in a total internal reflection process. The penetration depth of the evanescent wave ranges typically from 0.5-2 μm, well below the thickness of the hair cuticle. Therefore, the infrared spectra measured in the ATR mode with the transverse incidence are primarily dominated by the signal of the outmost cuticle layer. Destructive infrared measurements have been conducted on the hair with horizontal cuts, exhibiting noticeable spectral variation among different regions of the cross-section.[16–18] Hence, in order to acquire the overall infrared property of the whole hair structure, the transmission spectra on individual hair are directly probed without horizontal cuts, as illustrated in Fig. 1c. The high-quality spectrum from direct microscopic transmission allows for precise comparison of intrinsic spectral weight, which provides non-destructively tracing of the spectral evolution as hair grows.

In this work, we investigate the infrared transmission properties of human hair using FTIR microscopy. Prominent band filtering behaviors are observed in the spectra. The transmission spectra are well reproduced from the respective absorption of the hair cuticle, cortex, and medulla. The high spatial resolution of infrared micro-spectroscopy further offers the ability to compare the infrared spectrum at different regions of hair. In a case study of adult-onset Still's disease (AOSD), the corresponding infrared transmission displays a systematic variation of spectral weight. Although the patient population in this study is not enough for solid conclusion, it reflects a possible link between spectrum and disease-induced hair structure change. We further quantify the geometry effects of the hair's internal structure by means of the finite-element analysis method, and the results suggest that the variation of spectral weight probably correlates with the structure variation of the hair cortex. These findings point to the potential of the infrared micro-spectroscopy as a promising tool for non-destructive test for individual human hair.

**Results and discussion**

Fig. 1d displays the high-resolution transmission infrared micro-spectroscopy of a black hair sample obtained from a healthy Asian female. The infrared light transmits perpendicular through the horizontally placed hair sample, as illustrated in the inset of Fig. 1d. The optical slits of FTIR microscope were set to be slightly smaller than the hair diameter. In this configuration, the contributions from all three parts of the hair are accounted for, reproducing the light transmission in infrared therapy. The transmission spectrum shows strong band filtering behavior as a function of wavenumber ($\omega$), with low transmission in the range of 600-760 cm$^{-1}$, 1060-1690 cm$^{-1}$, and 2880-3500 cm$^{-1}$. Four prominent transmission peaks of 950 cm$^{-1}$, 1924 cm$^{-1}$, 2260 cm$^{-1}$, and 3810 cm$^{-1}$ are detected in the NIR and MIR regimes ranges of 580-4000 cm$^{-1}$ (2.5-17.2 μm in wavelength). The largest transmission is about 53.6% at 3810 cm$^{-1}$. Since the transmission peaks are not expected to match the chemical absorption peaks, the generation and variation of these peaks in the transmission spectrum above are likely to be closely related to the structural properties of the sample. These peaks of the transmission spectrum are repeatable and much larger than the noise, permitting them as features rather than artifacts. For comparison, in Fig. 1e, we present the absorption spectra of hair cuticle, cortex, and medulla obtained from the ATR-FTIR measurement on the hair cross-section in previous studies[16–18]. The absorption spectra of these three parts show peaks with similar shapes and positions but different spectral weights. These low-transmission bands in the current experiment agree well with these absorption peaks, while the transmission peaks locate in the low-absorption regions.

In principle, the absorption and transmission spectra are equivalent and interconvertible for a given objective. However, for regions with strong absorption, the transmission signal is challenging to be precisely probed experimentally and is usually overwhelmed in the background. It causes the characteristic peaks in the low-transmission bands of the transmission spectrum to be absent, while the absorption peaks in the absorption spectrum become prominent. Likewise, for regions with strong transmission, the characteristic peak signals in absorption are quite difficult to precisely probed. Thus, the ATR and transmission modes of FTIR are sensitive to strong absorption (weak transmission) and strong transmission (weak absorption) regions, respectively. It clearly indicates that the transmission FTIR of human hair is capable of serving as an important complement to the extensive research on the infrared absorption by ATR-FTIR. For example, the peak intensity of typical black-body radiation spectral at 310 K (around normal human body temperature) is around 1068 cm$^{-1}$, just at the boundary of the second low-transmission band of 1060-1690 cm$^{-1}$ of hair shown in Fig. 1d. Therefore, in conditions of high-precision thermal radiation detection, such as temperature imaging of tumor, this band-filtering effect should be considered for regions covered with hair. Similarly, for conditions of high-precision light projection, such as infrared therapy, it is also necessary to avoid these low-transmission bands mentioned above.

The implementation of transverse transmission configuration further provides a rapid and non-destructive approach to investigate the spatial variation of infrared spectrum in hair by FTIR microscope. As illustrated in the inset of Fig. 2a, we take advantage of the high-spatial resolution of FTIR microscope and acquire the infrared spectrum within a short length (<30 μm) of the hair sample. The growth rate of hair is around 1 cm per month. It, therefore, may be used to track the whole body physiological change during a certain period, which is at least several months.[30] We intentionally avoid using long hair samples to ensure less external damage from daily activity. The hair sample was collected directly from the scalp without being stained, permed, or experiencing other types of severe physical damage during the growth. For each sample, the spectra were collected at the top, middle, and bottom regions to capture the variation over time. The infrared spectra shown in Fig.2a were obtained from two samples from the same host body (Asian female, 29 years old, presumed healthy in the past year). The two samples marked as H1 and H2 are about 5.9 cm and 6.8 cm in length, measuring from the hair follicle. Since slight variations might occur in diameter between different samples and thus cause a shift in absolute transmittance, we normalize the spectra with the peak height at 3810 cm$^{-1}$ and the spectra baseline in order to focus on the relative change of spectral weight. Without loss of generality, normalization with spectrum mapping to different ranges does not affect the subsequent analysis of the relative changes in spectral weight. After the normalization process, all six spectral curves overlap well with each other, exhibiting slight fluctuations (less than 2%) at specific peaks. It provides reasonably consistent evidence that the infrared transmission is a steady property for a given host as long as no notable change in physiology occurs.

In Figure 2b, a case study of hair infrared transmission experiments on a patient with AOSD is presented. AOSD is a rare multisystemic disorder that causes inflammatory conditions in multiple organs.[31–34] It is generally perceived as a complex (multigenic) autoinflammatory syndrome. The clinical manifestations of AOSD usually include high spiking fever, arthralgia, evanescent skin rash, neutrophilic leukocytosis, and abnormal liver function results.[32] Up to now, diagnosis of AOSD remains challenging due to the polymorphic clinical presentation and largely relies on the exclusion of other possible diseases.[32–34] Early diagnosis of AOSD may help to improve the prognosis[35] and therefore call for validated biomedical examination. Since the hair growth process is known to be affected by the appearance of inflammatory[36,37], we utilize FTIR microscope to track the change of infrared transmission. The hair samples were obtained from a 19-year-old Asia female (P-AOSD), who was diagnosed with AOSD. The patient exhibited symptoms of fever, rash, myalgia, and arthralgia for over four months before the hair samples were collected. The infrared transmission spectra at different regions were measured. The length of 0.3-6.5 cm represents the distance between the measured area and the hair follicle at the bottom. The transmission spectra in Fig. 2b are renormalized to Peak D (3810 cm$^{-1}$), as discussed above. A systematic increase of peak height with length (from red to blue line) is found at Peak A, B, and C. An increase of over 15% in transmittance is detected among different regions, as illustrated in Fig. 2b. Similar behaviors are also observed in another hair from the same patient. Taking a growth rate of 1 cm per month, the measured region at 6.5 cm corresponds to hair grown 6.5 months ago, within the onset time of AOSD. It is reasonable to expect that the change in transmittance correlates with the onset progress from six months ago (at 6.5 cm) to the present (at 0.3 cm), as indicated by the yellow arrow in Fig. 2b. Figure 2c shows the normalized infrared transmission spectra of a hair from the same AOSD patient collected approximately 11 months later. During this period, the patient experienced an asymptomatic state for over six months and developed slight symptoms of disease relapse right after the new batch of hair samples was collected. The yellow arrow in Fig. 2c indicates the increase and then decrease trend for Peak A, B, and C (from red to blue line), which occurred during the recovery and then the relapse of AOSD.

To further elaborate on the possible link between the change in hair infrared transmission spectra and AOSD, data from four hair samples of the same AOSD patient (P-AOSD) are collected and then plotted as a function of time in Fig. 3. In order to neutralize the diameter difference among different hairs, the relative transmittance signal for Peak A, B, and C are normalized to Peak D. The examined hair position is converted to time as estimated by the hair growth rate. Meanwhile, the evolution of AOSD in the patient was briefly labeled at the bottom of Fig. 3. From this, it is found that the relative transmittance of these peaks strongly decreases with (even slightly before) the onset or the relapse of AOSD. The peak transmittance remained at a low level during the illness and only gradually recovered back to normal in about three months after entering the asymptomatic state. These results reveal that the decrease of infrared transmittance in these peaks may be closely associated with AOSD.

It has to be acknowledged that the rarity of AOSD gives rise to a rare patient population and therefore the difficulty of replicating the experiment on a large scale. Thus the current study could not reach to the solid conclusion but reflects the potential of hair infrared transmission spectrum in tracing the variation retrospectively.

To better understand the observed spectral variation, we quantify the impact of the geometric size of hair structure and theoretically model the infrared transmission using the finite-element analysis method. As shown in Fig. 4a, a three-layer structure is constructed with $t_{Cu}$, $d_{Co}$, and $d_{Me}$ being the thickness of the cuticle, the diameter of the cortex, and the diameter of the medulla, respectively. The initial values of $d_{Me}$, $d_{Co}$, and $t_{Cu}$ are set as 8 μm, 50 μm, and 3 μm, respectively. The infrared absorption coefficient is simplified to be uniform within each part following the spectra data revealed in Fig. 1d. A beam of parallel light transmits through the hair from left to right as Fig. 4a depicts. The calculation of the transmission ratio $I/I_0$ at different optical wavenumbers ($\omega$) yields the transmission spectrum. Figures 4b and 4c display the two-dimensional spatial mappings of intensity profiles within the hair at two typical wavenumbers 1735 cm$^{-1}$ and 963 cm$^{-1}$. The intensity profile mapping suggests that the majority of the intensity loss locates in the middle cortex region, rather than cuticle and medulla. Combining the transmission data at different $\omega$ gives the transmission spectrum. As presented in Fig. 4d-f, we systematically investigate the evolution of hair transmission spectra as the geometric size of the cuticle, cortex, and medulla varies. The value of $t_{Cu}$, $d_{Co}$, and $d_{Me}$ vary around the initial values mentioned above, from 1 to 5 μm, 40 to 90 μm, and 0 to 10 μm, respectively. As $t_{Cu}$ increases from 1 μm to 5 μm ($d_{Co}$ and $d_{Me}$ fixed at 50 μm and 8 μm), the transmission peak at 1735 cm$^{-1}$ shows mild decline of 2.7%, and the transmission dip at 3265 cm$^{-1}$ remains almost the same value (Fig. 4d). When $d_{Co}$ increases from 40 μm to 90 μm ($t_{Cu}$ and $d_{Me}$ fixed at 3 μm and 8 μm), the transmission peak at 1735 cm$^{-1}$ and the dip at 3265 cm$^{-1}$ exhibit noticeable changes. The former presents a substantial drop of 19.8% (Fig. 4e), and the latter also decreases by about 5.5%. In contrast, the increase in $d_{Me}$ from 0 μm to 10 μm ($t_{Cu}$ and $d_{Co}$ fixed at 3 μm and 50 μm) only shifts the transmission spectra for less than 0.1% in the range of 620-3700 cm$^{-1}$, causing all the curves stacking together (Fig. 4f).

The variation of $t_{Cu}$, $d_{Co}$, and $d_{Me}$ are demonstrated in Figs. 4d-f has covered the typical geometric size range of hair structure. Hence, it is reasonable to believe that the transverse transmission property is mainly modulated by the cortex layer in our setup. The relative minor influence of the cuticle layer could result from the small thickness value, while the contribution from the hair medulla is negligible, probably owing to its small projection area in the normal plane of injected light. It is worth noting that the above simulation investigates the impact of the geometric size from the aspect of optics. The chemical- or biology-related mechanisms, such as chemical diffusion, bleaching, and host with tumor, as extensively discussed in literatures[18,20,21,26,29], may induce the change of infrared absorption as well. Most previous studies concentrated on the loss or change of hair medulla induced by these factors by measuring the spectra at hair cross-section[23–29], whereas the transverse transmission studied here mainly probes the structural change in the hair cortex.

As presented in Fig. 4, the infrared transmission property of hair is primarily controlled by the hair cortex. It differs from the previous studies focusing on the relationship between hair medulla and disease attacks.[23–29] On the other hand, the overall diameters of the hair samples were measured under the optical microscope, revealing no notable change with length since it is mainly determined by the size of the hair follicle. Therefore, the observed spectral evolution with length in Fig. 3 could be related to the absorption change of the cortex layer. The cortex is formed by spindle-shaped cells containing keratin proteins and structural lipids, which assemble parallel along the hair length. As discussed earlier, the infrared transmission peak refers to optical frequencies at which the measured sample shows the weakest absorption. Unlike the absorption peak, which is sensitive to the resonance frequency of chemical molecules, the transmission peak occurs at the frequency (or energy) that the sample's absorption is comparatively low. Hence, the increase in the height of the transmission peak corresponds to the decrease of a general optical attenuation coefficient without the resonance process. Typically, for a given material and a fixed size, the optical attenuation coefficient is higher in a more disordered structure. This is attributed to the fact that as the disorder increases, the optical scattering at the interface of internal units increases, thereby producing more considerable attenuation. In fact, the clinical presentation of AOSD does support the presence of more disordered hair in AOSD patients. As AOSD develops, the high spiking fever and multisystemic inflammatory[32] may happen from time to time, disturbing the regular growth and keratinization of the hair cortex. It may lead to a decrease in transmission coefficient, as observed in our experiment, as the AOSD evolves in time and results in a more disordered arrangement of cells and keratin in hair growth. We further simulate the evolution of transmission spectra in hairs with partial high-absorption cortex regions as shown in Fig. 5. As the area of high-absorption cortex regions (marked as light purple regions in Fig. 5a) increases from 0% to 50%, the peak intensity around 1735 cm$^{-1}$ systematically decreases for over 10%. These results strengthen the link between cortex's structure variations and transmission spectra and point out the potential that the transmission infrared micro-spectroscopy of hair may be used to help track the evolution of AOSD in the future. Further experiments on the evolution of protein structures may contribute to directly clarifying the underlying mechanism of spectrum change.

In conclusion, we propose that the transmission-mode microscopic infrared spectroscopy could serve as an efficient tool for testing individual human hair. According to the spatial-resolved infrared spectra, the hair from an AOSD patient shows substantial variation in infrared transmission as the disease evolves or recrudesces. Combining the experimental results and the finite-element analysis, we systematical analyze the geometric size effect of the hair cuticle, cortex, and medulla and find the infrared transmission is mainly dominated by the cortex layer. Our work not only generates fresh insight into a better understanding of human hair's infrared transmission spectrum, but also presents a possibility of tracing the pathological change during the hair growth retrospectively.

## Experimental
### Hair sample preparation
Hair samples used in this study were collected directly from the scalp of the host body and had not been stained, permed, or experienced other types of severe physical damage during the growth. The internal short hair samples were selected to avoid possible damage during the host's daily life. The samples were cleaned with ethanol to remove possible residue on the surface before the infrared transmission experiments.
### Transmission infrared micro-spectroscopy
The hair samples were characterized by Fourier transform infrared spectrometer (FTIR Bruker Vertex 80V) with infrared microscope (Bruker Hyperion 2000). To avoid the influence of water vapor and carbon dioxide, the spectrometer is evacuated, and the microscope is kept in a nitrogen atmosphere. The infrared beam was emitted by the globar and modulated by the spectrometer before entering the microscope through the KBr window. The beam passed through the sample adhered to the transmission holder with the signal detected by a mercury cadmium telluride detector. The experiments were performed at near and middle infrared regimes from 580 to 4000 $cm^{-1}$ in wavenumber (2.5 to 17.2 μm in wavelength). The magnification of the microscope is set as 36X. The acquisition time and spectral resolution for each spectrum were fixed at three minutes and 4 $cm^{-1}$, respectively.

**Finite-element analysis**

To simulate the transmission spectrum of human hair, the finite-element analysis was conducted using the COMSOL Multiphysics software. During the simulation, we simplified the structure to a coaxial cylinder with three uniform layers. From the inside out, the three layers are the medulla, cortex, and cuticle, respectively. In the infrared transmission model, we ignored the light scattering from each hair structure and set the monochromatic light beam perpendicular to the longitudinal axis. With these assumptions, we simulated the infrared transmission process in the two-dimensional cross-section using the Lambert-Beer law. For solid substances, Lambert-Beer law describes the relationship between the absorbed light intensity at a certain wavelength and the thickness of the substance. We write the Lambert-Beer law in differential form as,

$$\frac{\partial I}{\partial x} = -\alpha I,$$

where $I$ is the intensity of the light beam, $x$ is the coordinate along the propagation direction of the beam, and $\alpha$ is the absorbance coefficient of the transmitted substance. The absorbance data of these three parts was extracted from previous studies[16–18] with the ATR-FTIR measurement on the hair cross-section. Using the General Form PDE module of COMSOL, we calculated the total absorbance of the hair cross-section and converted it to transmission data.

The diameter of human hair is typically 50-100 μm. The medulla, with a diameter of 5-10 μm, is surrounded by the cortex. The latter comprises the hair bulk with a diameter in the range of 40 to 95 μm. For the outermost layer-cuticle, the thickness is usually less than 5 μm[18]. During the simulation, we varied the diameter of the medulla and cortex and the thickness of the cuticle to quantify the geometric size effect on the transmission spectrum. The absorbance of the light purple region in Fig. 5 is set to be four times of that of normal cortex.

**Acknowledgements**

C.L. was supported by the Youth Medical Talents-General Practitioner Program of Shanghai Medicine and Health Development Foundation. X.Y. was supported by the National Natural Science Foundation of China (grant no. 12174104, no. 62005079 and no. 62111530237), the Shanghai Sailing Program (grant no. 20YF1411700), the International Scientific and Technological Cooperation Project of Shanghai (grant no. 20520710900) and a start-up grant from East China Normal University. C.Z. was supported by the Shanghai Sailing Program (grant no. 20YF1402300) and a start-up grant from Fudan University. We thank Prof. W. Yu for help in COMSOL simulation.

**Competing interests**
The authors declare no competing interests.

**Ethics and Consent Statement**
The study was carried out under the ethical approval of Ethics Committee of Zhongshan Hospital, Fudan University (B2022-489). The patients provided written informed consent to perform all necessary investigations and to use them for research purposes and publication.

**Data Availability Statement**
The data that support the findings of this study are available from the corresponding author upon reasonable request.

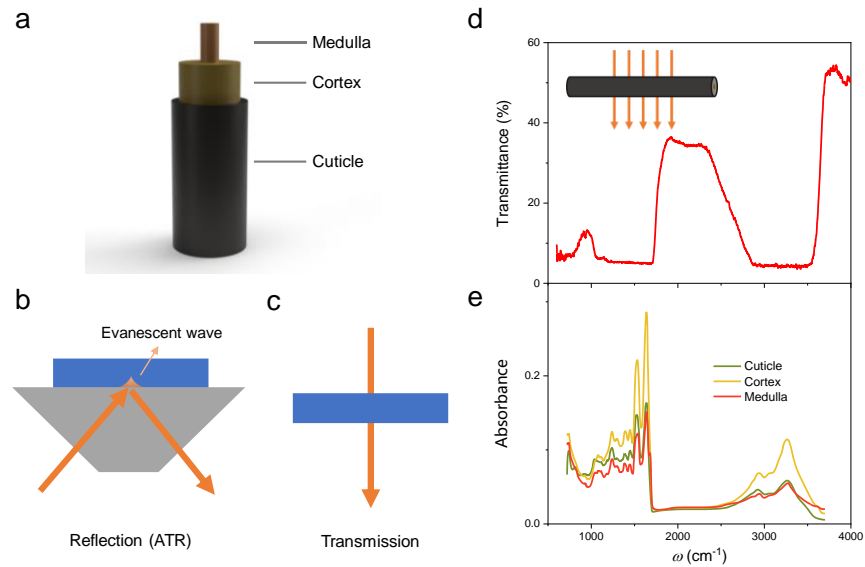

**Figure 1. Schematic illustration of the hair structure and different infrared spectral measurements. a**, the internal hair structure formed by three different parts: the inner medulla, the middle cortex layer, and the outmost cuticle layer. **b-c**, the ATR reflection mode (b) and the transmission mode (c) of FTIR experiments. The evanescent wave is generated in the total internal reflection of the ATR mode and penetrates into the sample (blue area) with a depth of 0.5-2 μm, which makes the ATR-FTIR a surface-sensitive technique. The transmission mode of FTIR directly measures the infrared beam transmitting through the sample, which probes the overall bulk property. Grey trapezoid denotes the ATR prisms. **d**, the transmission infrared micro-spectroscopic study on a hair sample. The inset is the measurement configuration. **e**, the absorption spectra of the hair cuticle, cortex, and medulla extracted from previous studies[16–18].

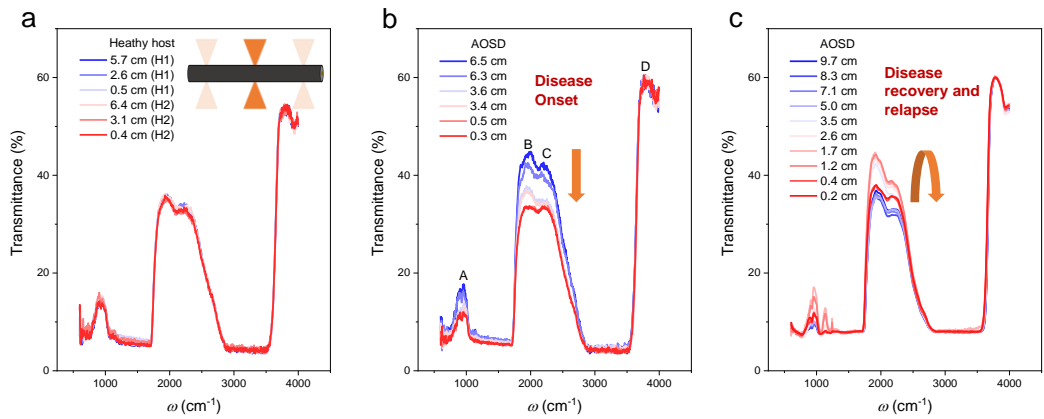

**Figure 2. A case study of hair infrared transmission experiments on a patient with AOSD.**
**a**, normalized infrared transmission spectra obtained from different regions of two samples from the same healthy host body. The inset illustrates the realization of spatial-resolved transmission infrared micro-spectra in hair. The length represents the distance between the measured area and the hair follicle. **b**, normalized infrared transmission spectra obtained from different regions of a hair from a patient with AOSD (P-AOSD). The yellow arrow indicates the decrease of transmittance for Peak A, B, and C, which occurs during the onset of the disease (from blue to red line). **c**, normalized infrared transmission spectra of a hair from the same AOSD patient collected about 11 months later. The yellow arrow indicates the increase and then decrease trend for Peak A, B, and C (from red to blue line), which occurred during the recovery and then relapse of AOSD.

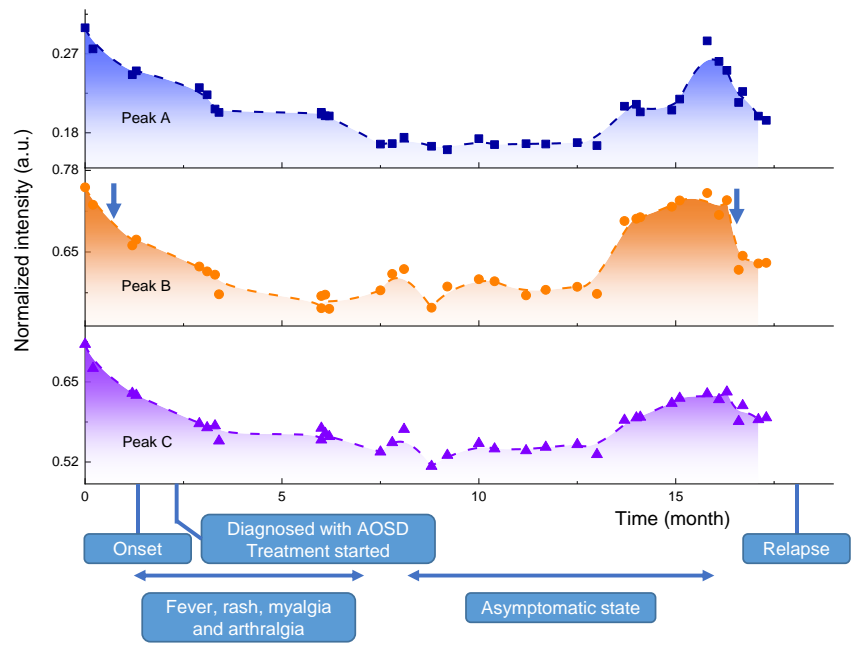

**Figure 3. Evolution of relative transmittance on the AOSD patient.** The relative transmittance data collected from four hair samples from the same AOSD patient (P-AOSD) is shown in time along with the evolution of the disease. Prominent decay in peak intensity (especially for Peak B as marked by the blue arrows) is observed slightly before the onset and the relapse of the disease.

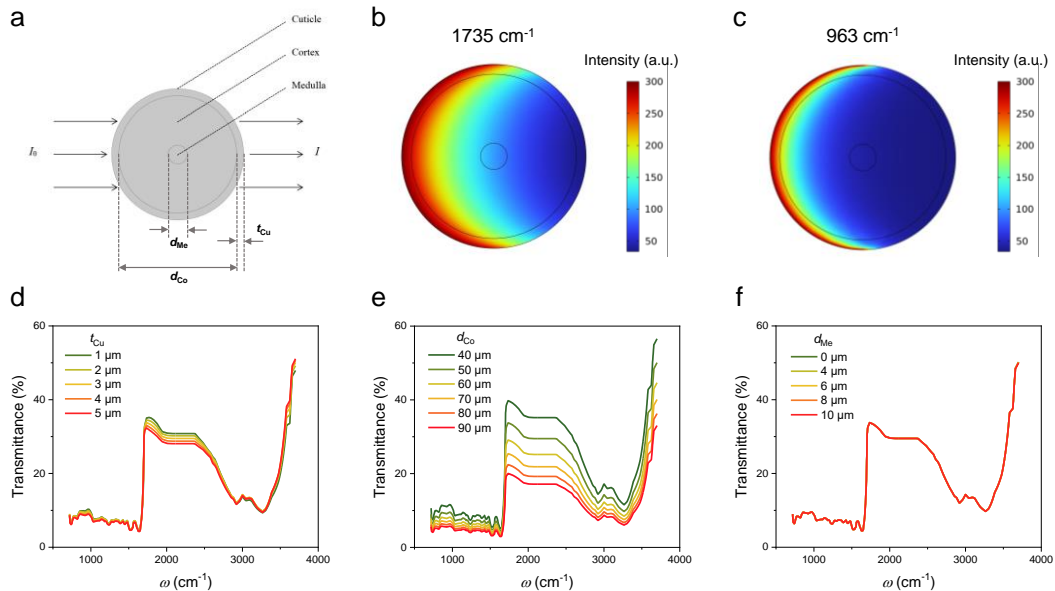

**Figure 4. The finite-element analysis of the influence of hair geometric size on the infrared transmission. a**, the hair infrared transmission model used in the simulation. **b-c**, the spatial mappings of light intensity profiles at two typical wavenumbers 1735 cm⁻¹ and 963 cm⁻¹. **d-f**, the evolution of transmission spectra as the geometric size of the cuticle, cortex, and medulla varies.

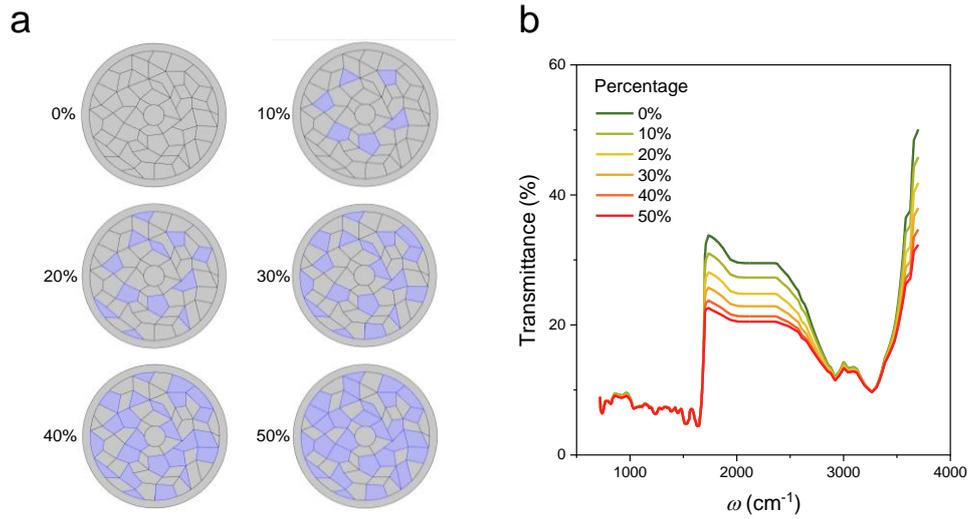

**Figure 5. Influence of high-absorption cortex region on the infrared transmission. a**, the illustration of different high-absorption cortex regions (light purple) in simulation. The marked value of 0% to 50% represents the area percentage of cortex regions with enhanced infrared absorption. **b**, the evolution of transmission spectra as the high-absorption cortex region increases.